\documentclass[conference]{IEEEtran}
\usepackage{cite}
\usepackage[dvips]{graphicx,color}
\usepackage[cmex10]{amsmath}
\usepackage{amssymb,enumerate}
\usepackage{environ,tikz}
\usepackage{flushend}

\makeatletter
\newsavebox{\measure@tikzpicture}
\NewEnviron{scaletikzpicturetowidth}[1]{%
  \def\tikz@width{#1}%
  \def\tikzscale{1}\begin{lrbox}{\measure@tikzpicture}%
  \BODY
  \end{lrbox}%
  \pgfmathparse{#1/\wd\measure@tikzpicture}%
  \edef\tikzscale{\pgfmathresult}%
  \BODY
}
\makeatother

\newcommand{\Prb}{\mathsf{P}}\newcommand{\Exp}{\mathsf{E}}
\newcommand{\Lpl}{\mathcal{L}}
\newcommand{\dd}{\mathrm{d}}\newcommand{\ee}{\mathrm{e}}
\newcommand{\C}{\mathbb{C}}\newcommand{\R}{\mathbb{R}}
\newcommand{\N}{\mathbb{N}}\newcommand{\Z}{\mathbb{Z}}
\newcommand{\B}{\mathcal{B}}

\newcommand{\SIR}{\mathsf{SIR}}
\newcommand{\Gam}{\mathrm{Gam}}
\newcommand{\Cell}{\mathcal{C}}

\newcommand{\ind}[1]{\boldsymbol{1}_{#1}}
\newcommand{\eqd}{\mathrel{\buildrel d \over =}}

\let\Bar\overline

\newtheorem{proposition}{Proposition}
\newtheorem{theorem}{Theorem}
\newtheorem{corollary}{Corollary}
\newtheorem{lemma}{Lemma}
\newtheorem{remark}{Remark}

\begin{document}\sloppy\allowbreak\allowdisplaybreaks

\title{A Sufficient Condition for Tail Asymptotics of SIR Distribution
  in Downlink Cellular Networks}
\author{%
  \IEEEauthorblockN{Naoto Miyoshi}
  \IEEEauthorblockA{Department of Mathematical and Computing Science\\
    Tokyo Institute of Technology\\
    2-12-1-W8-52 Ookayama, Tokyo 152-8552, Japan\\
    Email: miyoshi@c.titech.ac.jp}
  \and
  \IEEEauthorblockN{Tomoyuki Shirai}
  \IEEEauthorblockA{Institute of Mathematics for Industry\\
    Kyushu University\\
    744 Motooka, Fukuoka 819-0395, Japan\\
    Email: shirai@imi.kyushu-u.ac.jp}%
}
    
\maketitle

\begin{abstract}
We consider the spatial stochastic model of single-tier downlink
cellular networks, where the wireless base stations are deployed
according to a general stationary point process on the Euclidean plane
with general i.i.d.\ propagation effects.
Recently, Ganti \& Haenggi (2016) consider the same general cellular
network model and, as one of many significant results, derive the tail
asymptotics of the signal-to-interference ratio (SIR) distribution.
However, they do not mention any conditions under which the result
holds.
In this paper, we compensate their result for the lack of the
condition and expose a sufficient condition for the asymptotic result
to be valid.
We further illustrate some examples satisfying such a sufficient
condition and indicate the corresponding asymptotic results for the
example models.
We give also a simple counterexample violating the sufficient
condition.
\end{abstract}

\section{Introduction}

In the design and analysis of wireless communication networks, the
signal-to-interference ratio (SIR), defined in the next section, is a
key quantity.
In this paper, we consider the probability distribution of the SIR in
the spatial stochastic models of downlink cellular networks, where the
wireless base stations (BSs) are deployed according to spatial point
processes on the Euclidian plane (see, e.g.,
\cite{BaccBlas09a,Haen13,Mukh14}).
The SIR distribution in these cellular network models can be provided
as a closed-form or a numerically computable form for some restricted
cases such as the BSs are deployed according to homogeneous Poisson
point processes or kinds of determinantal point processes with
specific propagation effects of fading and shadowing (see, e.g.,
\cite{AndrBaccGant11,MiyoShir14a}).
However, such special cases can not always represent the actual BS
deployments and propagation effects in the real cellular networks, so
that some approximation and/or asymptotic approaches have been
proposed to cope with more general models (see, e.g.,
\cite{NagaMiyoShir14,Haen14,GuoHaen15,GantHaen15}).

In the current paper, we focus on the tail asymptotics of the SIR
distribution in the single-tier network models, where the BSs are
deployed according to general stationary point processes with general
propagation effects.
Recently, Ganti \& Haenggi~\cite{GantHaen15} consider the same general
cellular network models and investigate the asymptotics of the SIR
distribution both at the origin and at infinity.
In~\cite{GantHaen15}, they derive the tail asymptotic result which is
just our concern, but they do not mention any conditions under which
the result holds.
In their proof, a technique of changing the order of the limit and
integrals is used, which generally requires a kind of uniform
integrability condition.
This paper then compensates \cite{GantHaen15} for the lack of the
uniform integrability condition and exposes a sufficient condition for
the order change of the limit and integrals.
We further give some examples satisfying such a sufficient condition
as well as a counterexample violating it.

The rest of the paper is organized as follows:
First, we describe the spatial stochastic model of single-tier
downlink cellular networks and define the SIR for the typical user in
the next section.
Section \ref{sec:asymptotic_result} states the main result, where we
give a sufficient condition under which the tail asymptotics of the
SIR distribution is properly obtained. In section~\ref{sec:examples},
we illustrate some examples satisfying the condition and indicate the
asymptotic results for the corresponding models of the examples.
We further give a simple counterexample violating the   sufficient
condition.

\section{Network model}

Let $\Phi=\{X_i\}_{i\in\N}$ denote a point process on $\R^2$\!, which
is assumed to be ordered such that $|X_1| \le |X_2| \le \cdots$.
Each point~$X_i$, $i\in\N$, represents the location of a BS of the
cellular network and we refer to the BS located at $X_i$ as BS~$i$.
We assume that the point process~$\Phi$ is simple and locally finite
almost surely in $\Prb$ ($\Prb$-a.s.) and also stationary with
intensity~$\lambda\in(0,\infty)$.
Assuming further that all BSs transmit at the same power level and
each user is associated with the nearest BS, we focus on a typical
user located at the origin~$o=(0,0)$.
Let $H_i$, $i\in\N$, denote the random propagation effect representing
the fading and shadowing from BS~$i$ to the typical user, where $H_i$,
$i\in\N$, are mutually independent and identically distributed
(i.i.d.), and independent of the point process~$\Phi$ as well.
The path-loss function representing the attenuation of signals with
distance is given by $\ell(r)=r^{-2\,\beta}$, $r>0$, for $\beta>1$.
The downlink SIR for the typical user is then given by
\begin{equation}\label{eq:SIR}
  \SIR_o = \frac{H_1\,\ell(|X_1|)}
                {\sum_{i=2}^\infty H_i\,\ell(|X_i|)},
\end{equation}
where we should recall that $X_1$ is the nearest point of $\Phi$ from
the origin.

Our concern in the current paper is the tail asymptotics of the SIR
distribution; that is, the asymptotic behavior of
$\Prb(\SIR_o>\theta)$ as $\theta\to\infty$.

\section{General asymptotic result}\label{sec:asymptotic_result}

In this and later sections, $\Prb^0$ and $\Exp^0$ denote respectively
the Palm probability and the corresponding expectation with respect to
the marked point process $\Phi_H = \{(X_i, H_i)\}_{i\in\N}$ (see,
e.g., \cite[Sec.~1.4]{BaccBlas09a}).
Note that $\Prb^0(\Phi(\{o\})=1)=1$ while $\Prb(\Phi(\{o\})\ge1)=0$.
Note also that, due to the mutual independence of
$\Phi=\{X_i\}_{i\in\N}$ and $\{H_i\}_{i\in\N}$, $\Prb^0(H_i\in C) =
\Prb(H_i\in C)$ for $C\in\B(\R_+)$.
When we consider $\Phi$ under the Palm distribution $\Prb^0$, we use
the index~$0$ for the point at the origin; that is, $X_0=o=(0,0)$.

To give the main theorem which is a refinement of Theorem~4 of
\cite{GantHaen15}, we first define the typical Voronoi cell and its
circumscribed radius.
For a point process~$\Phi$ on $\R^2$ and a point~$X_i$ of $\Phi$, the
Voronoi cell of $X_i$ with respect to $\Phi$ is defined as the set;
\[
  \Cell(X_i)
  = \{x\in\R^2: |x-X_i| \le |x-X_j|, X_j\in\Phi\};
\]
that is, the set of points in $\R^2$ whose distance to $X_i$ is not
greater than that to any other points of $\Phi$.
The typical Voronoi cell is then $\Cell(o)$ under the Palm
distribution~$\Prb^0$ and its circumscribed radius, denoted by $R(o)$,
is the radius of the smallest disk centered at the origin and
containing $\Cell(o)$ under $\Prb^0$.

\begin{theorem}\label{thm:general}
We suppose the following.
\begin{enumerate}[(A)]
\item\label{condA} For the point process~$\Phi=\{X_i\}_{i\in\N}$,
  $\Exp^0(R(o)^2)<\infty$ and $\Exp^0(|X_k|^2)<\infty$ for any
  $k\in\N$.
\item\label{condB} For the sequence of the propagation effects
  $\{H_i\}_{i\in\N}$, $\Exp({H_1}^{1/\beta})<\infty$ and there exist
  $\alpha>0$ and $c_H>0$ such that the Laplace transform $\Lpl_H$ of
  $H_i$, $i\in\N$, satisfies $\Lpl_H(s)\le c_H\,s^{-\alpha}$ for
  $s\ge1$.
\end{enumerate}
It then holds that
\begin{align}\label{eq:GH15Thm4}
 &\lim_{\theta\to\infty}\theta^{1/\beta}\,
    \Prb(\SIR_o > \theta)
 \nonumber\\
 &= \pi\,\lambda\,
    \Exp({H_1}^{1/\beta})\,
    \Exp^0\biggl[
      \biggl(
        \sum_{i\in\N} \frac{H_i}{|X_i|^{2\,\beta}}
      \biggr)^{-1/\beta}
    \biggr].
\end{align}
\end{theorem}

One can show that the right-hand side of \eqref{eq:GH15Thm4} does not
depend on the intensity $\lambda$ of the point process $\Phi$ (see the
remark of Definition~4 in \cite{GantHaen15}).

\begin{remark}
The right-hand side of \eqref{eq:GH15Thm4} in
Theorem~\ref{thm:general} is identical to $\mathsf{EFIR}^\delta$ in
Theorem~4 of \cite{GantHaen15}; that is, that theorem and our
Theorem~\ref{thm:general} assert the same result.
The difference between the two theorems is that we provide a set of
conditions~(\ref{condA}) and (\ref{condB}), the role of which is
discussed in the proof and the remarks thereafter.
\end{remark}

\begin{IEEEproof}
Let $F_H$ denote the distribution function of $H_i$, $i\in\N$, and let
$\Bar{F_H}(x) = 1-F_H(x)$.
By~\eqref{eq:SIR} and $\ell(r)=r^{-2\beta}$, $r>0$, the tail
probability of the downlink SIR for the typical user is expressed as
\begin{equation}\label{eq:BarF_H}
  \Prb(\SIR_o > \theta)
  = \Exp\Bar{F_H}\biggl(
      \theta\,|X_1|^{2\,\beta}
      \sum_{i=2}^\infty
        \frac{H_i}{|X_i|^{2\,\beta}}
    \biggr).
\end{equation}
Applying the Palm inversion formula~(see, e.g.,
\cite[Sec.~4.2]{BaccBlas09a}) to the right-hand side above,
\begin{align*}
  &\Prb(\SIR_o > \theta)
  \\
  &= \lambda\int_{\R^2}
       \Exp^0\biggl[
         \Bar{F_H}\biggl(
           \theta\, |x|^{2\,\beta}
           \sum_{i\in\N}\frac{H_i}{|X_i-x|^{2\,\beta}}
         \biggr)\,
         \ind{\Cell(o)}(x)
       \biggr]\,
     \dd x
  \\
  &= \theta^{-1/\beta}\,\lambda
     \int_{\R^2}
       \Exp^0\biggl[
         \Bar{F_H}\biggl(
           |y|^{2\,\beta}
           \sum_{i\in\N}
             \frac{H_i}
                  {|X_i-\theta^{-1/(2\,\beta)}\,y|^{2\,\beta}}
         \biggr)
  \\
  &\qquad\qquad\qquad\qquad\mbox{}\times
         \ind{\Cell(o)}(\theta^{-1/(2\,\beta)}\,y)
       \biggr]\,
     \dd y,
\end{align*}
where the second equality follows by substituting
$y=\theta^{1/(2\,\beta)}\,x$.
Here, if we can find a random function $A$ satisfying
\begin{gather}
  \Bar{F_H}\biggl(
    |y|^{2\,\beta}
    \sum_{i\in\N}
      \frac{H_i}
           {|X_i-\theta^{-1/(2\,\beta)}\,y|^{2\,\beta}}
  \biggr)\,
  \ind{\Cell(o)}(\theta^{-1/(2\,\beta)}\,y)
  \nonumber\\
  \le A(y),
  \quad\text{$\Prb^0$-a.s.,}
  \label{eq:A1}\\
  \int_{\R^2} \Exp^0 A(y)\,\dd y < \infty,
  \label{eq:A2}
\end{gather}
the dominated convergence theorem leads to
\begin{align}\label{eq:limit}
 &\lim_{\theta\to\infty}
    \theta^{1/\beta}\,\Prb(\SIR_o>\theta)
 \nonumber\\
 &= \lambda\int_{\R^2}
      \Exp^0\Bar{F_H}\biggl(
        |y|^{2\,\beta}
        \sum_{i\in\N}\frac{H_i}
                         {|X_i|^{2\,\beta}}
      \biggr)\,
    \dd y.
\end{align}
We leave finding such an $A$ and approve \eqref{eq:limit} for a
moment.
Substituting
$z=\bigl(\sum_{i\in\N}H_i/|X_i|^{2\,\beta}\bigr)^{1/(2\,\beta)}\,y$ to
the last integral in \eqref{eq:limit} yields
\begin{align}\label{eq:integral1}
 &\int_{\R^2}
    \Exp^0\Bar{F_H}\biggl(
      |y|^{2\,\beta}
      \sum_{i\in\N} \frac{H_i}{|X_i|^{2\,\beta}}
    \biggr)\,
  \dd y
 \nonumber\\
 &= \Exp^0\biggl[
      \biggl(
        \sum_{i\in\N} \frac{H_i}{|X_i|^{2\,\beta}}
      \biggr)^{-1/\beta}
    \biggr]
    \int_{\R^2}\Bar{F_H}(|z|^{2\,\beta})\,\dd z,
\end{align}
and furthermore,
\begin{align}\label{eq:integral2}
  \int_{\R^2}\Bar{F_H}(|z|^{2\,\beta})\,\dd z
  &= 2\,\pi\int_0^\infty\Bar{F_H}(r^{2\,\beta})\,r\,\dd r
  \nonumber\\
  &= \frac{\pi}{\beta}\,
     \int_0^\infty \Bar{F_H}(s)\,s^{1/\beta-1}\,\dd s
  \nonumber\\
  &= \pi\,\Exp({H_1}^{1/\beta}).
\end{align}
Hence, applying \eqref{eq:integral1} and \eqref{eq:integral2} to
\eqref{eq:limit}, we obtain \eqref{eq:GH15Thm4}.

It remains to find a function~$A$ satisfying \eqref{eq:A1} and
\eqref{eq:A2}.
Since $\Bar{F_H}$ is nonincreasing and $|X_i-y|\le |X_i|+R(o)$
$\Prb^0$-a.s.\ for $y\in\Cell(o)$, we can set the function~$A$
satisfying \eqref{eq:A1} as
\[
  A(y) =
  \Bar{F_H}\biggl(
    |y|^{2\,\beta}\,
    \sum_{i\in\N}
      \frac{H_i}
           {\bigl(|X_i| + R(o)\bigr)^{2\,\beta}}
  \biggr).
\]
Then, substituting $z = \bigl( \sum_{i\in\N} H_i / \bigl(|X_i| +
R(o)\bigr)^{2\,\beta} \bigr)^{1/(2\,\beta)}\,y$ and applying
\eqref{eq:integral2} again, we have
\begin{align*}
 &\int_{\R^2}\Exp^0 A(y)\,\dd y
 \\
 &= \pi\,\Exp({H_1}^{1/\beta})\,
    \Exp^0\biggl[
      \biggl(
        \sum_{i\in\N}
          \frac{H_i}{\bigl(|X_i| + R(o)\bigr)^{2\,\beta}}
      \biggr)^{-1/\beta}
    \biggr].
\end{align*}
For the second expectation on the right-hand side above, applying the
identity $x^{-1/\beta} = \Gamma(1/\beta)^{-1}\int_0^\infty
\ee^{-x\,s}\, s^{-1+1/\beta}\,\dd s$ yields
\begin{align*}
  &\Exp^0\biggl[
     \biggl(
       \sum_{i\in\N}
         \frac{H_i}{\bigl(|X_i| + R(o)\bigr)^{2\,\beta}}
     \biggr)^{-1/\beta}
   \biggr]
  \\
  &= \frac{1}{\Gamma(1/\beta)}
     \int_0^\infty
       s^{-1+1/\beta}\\
  &\qquad\qquad\mbox{}\times
       \Exp^0\biggl[
         \exp\biggl(
           -s \sum_{i\in\N}
                \frac{H_i}{\bigl(|X_i| + R(o)\bigr)^{2\,\beta}}
         \biggr)
       \biggr]\,
     \dd s
  \\
  &= \frac{1}{\Gamma(1/\beta)}
     \int_0^\infty
       s^{-1+1/\beta}\\
  &\qquad\qquad\mbox{}\times
       \Exp^0\biggl[
         \prod_{i\in\N}
           \Lpl_H\biggl(
             \frac{s}{\bigl(|X_i| + R(o)\bigr)^{2\,\beta}}
           \biggr)
       \biggr]\,
     \dd s,
\end{align*} 
where $\Gamma$ denotes Euler's Gamma function.
Recall that $X_i$, $i\in\N$, are ordered such that
$|X_1|<|X_2|<\cdots$.
Thus, by taking $k\in\N$ such that $\alpha\,\beta\,k>1$, and applying
$\Lpl_H(s)\le c_H\,s^{-\alpha}$ for $s\ge1$ from
condition~(\ref{condB}), we have
\begin{align}\label{eq:A_bound}
  &\int_0^\infty
     s^{-1+1/\beta}\,
     \Exp^0\biggl[
       \prod_{i\in\N}
         \Lpl_H\biggl(
           \frac{s}{\bigl(|X_i| + R(o)\bigr)^{2\,\beta}}
         \biggr)
     \biggr]\,
   \dd s
  \nonumber\\
  &\le \int_0^\infty
         s^{-1+1/\beta}\,
         \Exp^0\biggl[
           \Lpl_H\biggl(
             \frac{s}{\bigl(|X_k| + R(o)\bigr)^{2\,\beta}}
           \biggr)^k
         \biggr]\,
       \dd s
  \nonumber\\
  &\le \Exp^0\biggl[
         \int_0^{(|X_k| + R(o))^{2\,\beta}}
           s^{-1+1/\beta}\,
         \dd s
       \biggr]
  \nonumber\\
  &\quad\mbox{}
       + {c_H}^k\,
         \Exp^0\biggl[
           \bigl(|X_k| + R(o)\bigr)^{2\,\alpha\,\beta\,k}
  \nonumber\\
  &\qquad\quad\qquad\mbox{}\times
           \int_{(|X_k| + R(o))^{2\,\beta}}^\infty
             s^{-1+1/\beta-\alpha\,k}\,
           \dd s
         \biggr]
  \nonumber\\
  &= \beta\,
     \Bigl(
       1 + \frac{{c_H}^k}{\alpha\,\beta\,k-1}
     \Bigr)\,
     \Exp^0\bigl[
       \bigl(|X_k| + R(o)\bigr)^2
     \bigr].
\end{align}
Hence, the inequality $(a+b)^2 \le 2\,(a^2 + b^2)$ ensures
\eqref{eq:A2} under condition~(\ref{condA}) of the theorem.
\end{IEEEproof}

\begin{remark}
The differences between the proof of \cite{GantHaen15} and ours are as
follows.
The first and less essential one is that, in~\cite{GantHaen15}, they
modify the right-hand side of \eqref{eq:BarF_H} into an appropriate
form and then apply the Campbell-Mecke formula.
On the other hand, we apply the Palm inversion formula directly.
Second, \cite{GantHaen15} does not specify any condition under which
the result holds.
However, equality~\eqref{eq:limit} requires some kind of uniform
integrability condition to change the order of the limit and
integrals.
Our set of conditions~(\ref{condA}) and (\ref{condB}) in
Theorem~\ref{thm:general} gives a sufficient condition for this order
change to be valid.
\end{remark}

\begin{remark}
The condition~(\ref{condB}) claims that the Laplace transform of
$H_i$, $i\in\N$, decays faster than or equal to the power law.
Though this condition excludes the distribution with a mass at the
origin, it covers many practical distributions.
For example, the Gamma distribution~$\Gam(\alpha, b)$, $\alpha>0$,
$b>0$, has the Laplace transform~$\Lpl_H(s)=(1+b\,s)^{-\alpha}$ and we
can take $c_H\ge b^{-\alpha}$.
\end{remark}

The asymptotic constant of~\eqref{eq:GH15Thm4} in
Theorem~\ref{thm:general} depends on the point process~$\Phi$ and the
distribution~$F_H$ of the propagation effects.
The following proposition gives the impact of the propagation effects
to the asymptotic constant by comparing with the case without
propagation effects.

\begin{proposition}\label{thm:inequality}
Let $C(\beta, F_H)$ denote the limit on the right-hand side of
\eqref{eq:GH15Thm4}, specifying the dependence on the value of $\beta$
and the propagation effect distribution~$F_H$.
When $\Exp H_1 <\infty$, we have
\begin{equation}\label{eq:AsymIneq}
  C(\beta, F_H)
  \ge \frac{\Exp({H_1}^{1/\beta})}{(\Exp H_1)^{1/\beta}}\,
      C(\beta, \delta_1),
\end{equation}
where $\delta_1$ denotes the Dirac measure with the mass at the unity.
\end{proposition}

\begin{IEEEproof}
The result immediately follows from Jensen's inequality conditioned on
$\Phi=\{X_i\}_{i\in\N}$.
On the right hand-side of \eqref{eq:GH15Thm4}, since
$f(x)=x^{-1/\beta}$ is convex for $x>0$,
\begin{align*}
 &\Exp^0\biggl[
    \biggl(
      \sum_{i\in\N}\frac{H_i}{|X_i|^{2\,\beta}}
    \biggr)^{-1/\beta}
  \biggr]
  \ge \Exp^0\biggl[
         \biggl(
           \sum_{i\in\N}\frac{\Exp H_1}{|X_i|^{2\,\beta}}
         \biggr)^{-1/\beta}
       \biggr]
  \\
  &= \frac{1}{(\Exp H_1)^{1/\beta}}\,
     \Exp^0\biggl[
       \biggl(
         \sum_{i\in\N} \frac{1}{|X_i|^{2\,\beta}}
       \biggr)^{-1/\beta}
     \biggr],
\end{align*}
and \eqref{eq:AsymIneq} holds.
\end{IEEEproof}

\begin{remark}
When $F_H=\mathrm{Exp}(1)$ (Rayleigh fading without shadowing), the
result of Proposition~\ref{thm:inequality} coincides with the second
part of Theorem~2 in~\cite{MiyoShir15}.
In the inequality~\eqref{eq:AsymIneq}, it is easy to see (by Jensen's
inequality) that the coefficient $\Exp({H_1}^{1/\beta})/(\Exp
H_1)^{1/\beta}$ is smaller than or equal to the unity.
Now, suppose that $\Exp H_1 = 1$.
Then, the dominated convergence theorem (due to ${H_1}^{1/\beta}\le 1
+ H_1$ a.s.) leads to $\Exp({H_1}^{1/\beta})\to1$ as both
$\beta\downarrow1$ and $\beta\uparrow\infty$, which implies that
$C(\beta, F_H) \ge C(\beta, \delta_1)$ might be true when the value of
$\beta$ is close to the unity or sufficiently large.
\end{remark}

\section{Examples}\label{sec:examples}

\subsection{Poisson process networks}

In this section, we consider the homogeneous Poisson point
process~$\Phi$ with finite and nonzero intensity~$\lambda$.
We first confirm that $\Phi$ satisfies the condition~(\ref{condA}) of
Theorem~\ref{thm:general}.

\begin{lemma}\label{lem:Poi_A}
Let $\Phi=\{X_i\}_{i\in\N}$ denote the homogeneous Poisson point process
with intensity~$\lambda\in(0,\infty)$, where $X_i$, $i\in\N$, are
ordered such that $|X_1|<|X_2|<\cdots$.
Then, for $\epsilon>0$,
\begin{align}
  \Exp^0\ee^{\epsilon|X_k|} &<\infty,
  \quad k\in\N,
  \label{eq:Poi_distance}\\
  \Exp^0\ee^{\epsilon R(o)} &<\infty.
  \label{eq:Poi_radius}
\end{align}
\end{lemma}

This lemma ensures that both $|X_k|$, $k\in\N$, and $R(o)$ have any
order of moments.

\begin{IEEEproof}
Let $D_r$ denote the disk centered at the origin with radius~$r>0$.
By Slivnyak's theorem~(see, e.g., \cite[Sec.~1.4]{BaccBlas09a},
\cite{Sliv62}),
\begin{align*}
  \Prb^0(|X_k| > r)
  &= \Prb(|X_k| > r)
   = \Prb(\Phi(D_r) < k)\\
  &= \ee^{-\lambda\pi r^2}
     \sum_{j=0}^{k-1} \frac{(\lambda\,\pi\,r^2)^j}{j!}.
\end{align*}
Therefore, exploiting $\pi\lambda r^2 - \epsilon r \ge \pi\lambda
r^2/2$ for $r \ge 2\epsilon/(\pi\lambda)$,
\begin{align*}
  \Exp^0\ee^{\epsilon|X_k|}
  &= \int_0^\infty
       \ee^{\epsilon r}\,
       \frac{2\pi\lambda r\,(\pi\lambda r^2)^{k-1}}{(k-1)!}\,
       \ee^{-\pi\lambda r^2}\,
     \dd r
  \\
  &\le \ee^{2\epsilon^2/(\pi\lambda)}
       \int_0^{2\epsilon/(\pi\lambda)}
         \frac{2\pi\lambda r\,(\pi\lambda r^2)^{k-1}}{(k-1)!}\,
         \ee^{-\pi\lambda r^2}\,
       \dd r
  \\
  &\quad\mbox{}
     + \int_{2\epsilon/(\pi\lambda)}^\infty
         \frac{2\pi\lambda r\,(\pi\lambda r^2)^{k-1}}{(k-1)!}\,
         \ee^{-\pi\lambda r^2/2}\,
       \dd r
  \\
  &= \ee^{2\epsilon^2/(\pi\lambda)}\,
     P\Bigl(k, \frac{4\epsilon^2}{\pi\lambda}\Bigr)
     + 2^k\,Q\Bigl(k, \frac{2\epsilon^2}{\pi\lambda}\Bigr),
\end{align*}
and \eqref{eq:Poi_distance} holds, where $P$ and $Q$ denote
respectively the regularized lower and upper incomplete Gamma
functions.

On the other hand, for the circumscribed radius~$R(o)$ of the typical
Voronoi cell of $\Phi$, Calka~\cite[Theorem~3]{Calk02} shows that
\[
  \Prb^0(R(o) > r)
  \le 4\,\pi\,\lambda\,r^2\,\ee^{-\pi\lambda r^2}
  \quad\text{for $r\ge r_0\approx0.337$,}
\]
so that, we have
\begin{align*}
  \Exp^0\ee^{\epsilon R(o)}
  &= 1
   + \epsilon
     \int_0^\infty
       \ee^{\epsilon r}\,\Prb^0(R(o) > r)\,
     \dd r
  \\
  &\le 1
     + \epsilon
       \int_0^{r_0\vee(2\epsilon/(\pi\lambda))}
         \ee^{\epsilon r}\,
       \dd r
  \\
  &\quad\mbox{}
     + \epsilon
       \int_{r_0\vee(2\epsilon/(\pi\lambda))}^\infty
         4\,\pi\,\lambda\,r^2\,\ee^{-\pi\lambda r^2/2}\,
       \dd r
  \\
  &= \ee^{\epsilon r_0\vee(2\epsilon/(\pi\lambda))}
   + \frac{8}{(2\pi\lambda)^{1/2}}\,
     \Gamma\Bigl(
       \frac{3}{2}, 
       \frac{\pi\lambda{r_0}^2}{2}\vee\frac{2\epsilon^2}{\pi\lambda}
     \Bigr),
\end{align*}
and \eqref{eq:Poi_radius} holds, where $\Gamma(a,b)$ denotes the upper
incomplete Gamma function and $a\vee b=\max(a,b)$.
\end{IEEEproof}

We now apply Theorem~\ref{thm:general} and obtain the following.

\begin{corollary}\label{cor:Poisson}
Suppose that $\Phi=\{X_i\}_{i\in\N}$ is the homogeneous Poisson point
process.
When the propagation effect sequence~$\{H_i\}_{i\in\N}$ satisfies the
condition~(\ref{condB}) of Theorem~\ref{thm:general}, the right-hand
side of \eqref{eq:GH15Thm4} reduces to $(\beta/\pi)\,\sin(\pi/\beta)$.
\end{corollary}

\begin{IEEEproof}
Since the conditions of Theorem~\ref{thm:general} are fulfilled, the
result follows from the proof of Lemma~6 in \cite{GantHaen15}.
\end{IEEEproof}

\begin{remark}
The asymptotic result from Corollary~\ref{cor:Poisson} agrees with
that in \cite[Remark~4]{MiyoShir14a}, where the Rayleigh fading is
considered.
Corollary~\ref{cor:Poisson} states that the downlink coverage
probability in the Poisson cellular network is asymptotically
insensitive to the distribution of the propagation effects as far as
it satisfies the condition~(\ref{condB}) of Theorem~\ref{thm:general}.
\end{remark}

\subsection{Determinantal process networks}

In this section, we consider a general stationary and isotropic
determinantal point process~$\Phi$ on $\C\simeq\R^2$ with
intensity~$\lambda$.
Let $K$:~$\C^2\to\C$ denote the continuous kernel of $\Phi$ with
respect to the Lebesgue measure.
Then, the joint intensities (correlation functions)~$\rho_n$,
$n\in\N$, with respect to the Lebesgue measure are given by
\[
  \rho_n(z_1,z_2,\ldots,z_n)
  = \det\bigl(K(z_i, z_j)\bigr)_{i, j=1,2,\ldots,n},
\]
for $z_1$, $z_2$, \ldots, $z_n\in\C$.
Note that, due to the stationarity and isotropy, it holds that
$\rho_1(z)=K(z,z)=\lambda$ and that $\rho_2(0,z) = \lambda^2 -
|K(0,z)|^2$ depends only on $|z|$  for $z\in\C$.
In order for the point process~$\Phi$ to be well-defined, we assume
that (i)~the kernel~$K$ is Hermitian in the sense that $K(z,w) =
\Bar{K(w,z)}$ for $z$, $w\in\C$, where $\Bar{z}$ denotes the complex
conjugate of $z\in\C$, and (ii)~the integral operator on $L^2(\C)$
corresponding to $K$ is of locally trace class with the spectrum in
$[0,1]$; that is, for a compact set $C\in\B(\C)$, the restriction
$K_C$ of $K$ on $C$ has eigenvalues $\kappa_{C,i}$, $i\in\N$,
satisfying $\sum_{i\in\N}\kappa_{C,i}<\infty$ and
$\kappa_{C,i}\in[0,1]$ for each $i\in\N$ (see, e.g.,
\cite[Chap.~4]{HougKrisPereVira09}).

Concerning the condition~(\ref{condA}) of Theorem~\ref{thm:general},
we show the following.

\begin{lemma}\label{lem:determinantal}
\begin{enumerate}[(i)]
\item Let $X_i$, $i\in\N$, denote the points of $\Phi$ such that
  $|X_1|<|X_2|<\cdots$.
  Then, there exist $a_1>0$ and $a_2>0$ such that, for any $k\in\N$,
  we can take a sufficiently large $r>0$ satisfying
  \begin{equation}\label{eq:bound1}
    \Prb^0(|X_k| > r) \le a_1\,\ee^{-a_2\,r^2}\!.
  \end{equation}
\item Let $R(o)$ denote the circumscribed radius of the typical Voronoi
  cell~$\Cell(o)$ of $\Phi$.
  Then, there exist $b_1>0$ and $b_2>0$ such that, for $r>0$,
  \begin{equation}\label{eq:bound2}
    \Prb^0(R(o) > r) \le b_1\,\ee^{-b_2\,r^2}\!.
  \end{equation}
\end{enumerate}
\end{lemma} 

By Lemma~\ref{lem:determinantal}, it is easy to confirm, similar to
Lemma~\ref{lem:Poi_A}, that both $|X_k|$, $k\in\N$, and $R(o)$ have
any order of moments under $\Prb^0$.
To prove Lemma~\ref{lem:determinantal}, we use the following
supplementary lemma, the proof of which is in the appendix:

\begin{lemma}\label{lem:kernel}
The kernel $K$ of $\Phi$ satisfies
\begin{equation}\label{eq:boundM}
  \int_{\C} |K(0,z)|^2\,\dd z
  \le K(0,0)=\lambda.
\end{equation}
\end{lemma}

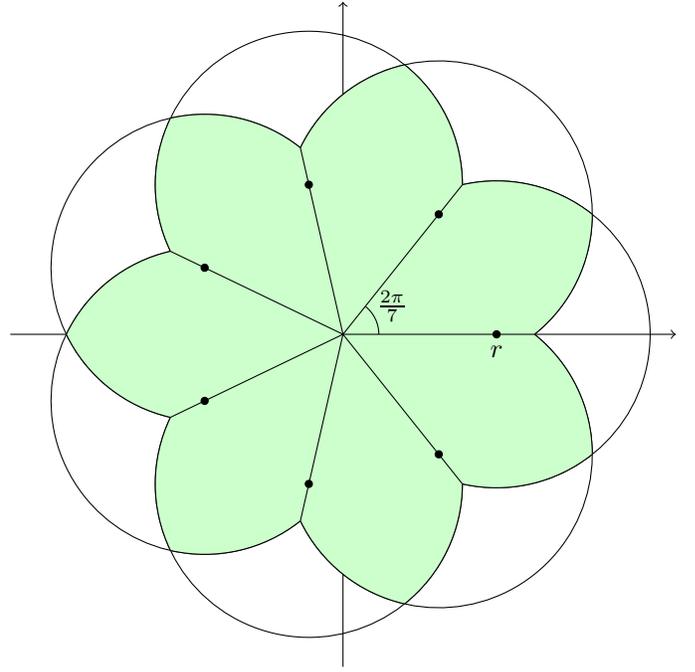
\begin{figure}
\begin{center}
\begin{scaletikzpicturetowidth}{\linewidth}
\begin{tikzpicture}[scale=\tikzscale]
\draw[->](-6.5,0)--(6.5,0);
\draw[->](0,-6.5)--(0,6.5);
\foreach \x in {0, 51.42857142857143, 102.8571428571429,
  154.2857142857143, 205.7142857142857, 257.1428571428571,
  308.5714285714286}
  \draw[rotate=\x] (3,0) circle (3cm);
\fill[green!20!white] (0,0) circle (3.74093881115cm);
\foreach \x in {0, 51.42857142857143, 102.8571428571429,
  154.2857142857143, 205.7142857142857, 257.1428571428571,
  308.5714285714286}
  \fill[rotate=\x,rotate around={51.4285714286:(3,0)},green!20!white]
  (3cm,0cm)--(6cm,0cm) arc (0:51.4285714286:3cm);
\foreach \x in {0, 51.42857142857143, 102.8571428571429,
  154.2857142857143, 205.7142857142857, 257.1428571428571,
  308.5714285714286}
  \fill[rotate=\x,rotate around={257.1428571428571:(3,0)},green!20!white]
    (3cm,0cm)--(6cm,0cm) arc (0:51.4285714286:3cm);
\foreach \x in {0, 51.42857142857143, 102.8571428571429,
  154.2857142857143, 205.7142857142857, 257.1428571428571,
  308.5714285714286}
  \draw[rotate=\x,rotate around={51.4285714286:(3,0)}]
  (6cm,0cm) arc (0:51.4285714286:3cm);
\foreach \x in {0, 51.42857142857143, 102.8571428571429,
  154.2857142857143, 205.7142857142857, 257.1428571428571,
  308.5714285714286}
  \draw[rotate=\x,rotate around={257.1428571428571:(3,0)}]
    (6cm,0cm) arc (0:51.4285714286:3cm);
\foreach \x in {0, 51.42857142857143, 102.8571428571429,
  154.2857142857143, 205.7142857142857, 257.1428571428571,
  308.5714285714286}
  \draw[rotate=\x](0,0)--(3.74093881115,0);
\foreach \x in {0, 51.42857142857143, 102.8571428571429,
  154.2857142857143, 205.7142857142857, 257.1428571428571,
  308.5714285714286}
  \filldraw[rotate=\x] (3,0) circle (.7mm);
\draw (3,0) node[below,inner sep=1.5mm]{$r$};
\draw (7mm,0mm) arc (0:51.4285714286:7mm) node[inner xsep=1.5mm,right]{$\frac{2\pi}{7}$};
\end{tikzpicture}
\end{scaletikzpicturetowidth}
\end{center}
\caption{Foss \& Zuyev's seven petals.}\label{fig:petals}
\end{figure}

\begin{IEEEproof}[Proof of Lemma~\ref{lem:determinantal}]
Let $\Prb^!$ denote the reduced Palm probability with respect to the
marked point process $\Phi_H=\{(X_i,H_i)\}_{i\in\N}$ and let $C$
denote a bounded set in $\B(\C)$.
Since the (reduced) Palm version of a determinantal point process is
also determinantal~(see, e.g., \cite{ShirTaka03}), $\Phi(C)$ under
$\Prb^!$ has the same distribution as $\sum_{i\in\N} B_{C,i}$ with
some kind of mutually independent Bernoulli random variables
$B_{C,i}$, $i\in\N$ (see, e.g., \cite[Sec.~4.5]{HougKrisPereVira09}).
Thus, the Chernoff-Hoeffding bound for an infinite sum with finite
mean~(see, e.g., \cite{Cher52,Hoef63} for a finite sum) implies that,
for any $\epsilon\in[0,1)$, there exists a
  $c_\epsilon>0$ such that
\begin{equation}\label{eq:C-H}
  \Prb^!\bigl(
    \Phi(C) \le \epsilon\,\Exp^!\Phi(C)
  \bigr)
  \le \ee^{-c_\epsilon\,\Exp^!\Phi(C)},
\end{equation}
where $\Exp^!$ denotes the expectation with respect to $\Prb^!$.
On the other hand, the kernel of the Palm version of $\Phi$ is given
by (see \cite{ShirTaka03})
\[
  K^0(z,w)
  = \frac{K(z,w)\,K(0,0) - K(z,0)\,K(0,w)}{K(0,0)},
  \;\; z,w\in\C.
\]
Therefore, the intensity function ($1$-correlation) of $\Phi$ under
$\Prb^!$ reduces to
\begin{equation}\label{eq:1-corr}
  \rho^0_1(z)
  = K^0(z,z)
  = \lambda - \frac{|K(0,z)|^2}{\lambda},
\end{equation}
so that, Lemma~\ref{lem:kernel} leads to
\begin{equation}\label{eq:Palmmean}
  \Exp^!\Phi(C)
  = \int_C \rho^0_1(z)\,\dd z
  \ge \lambda\,\mu(C) - 1,
\end{equation}
where $\mu$ denotes the Lebesgue measure on $\C$.

\paragraph*{\it Proof of (i)}
Note that $\Prb^0(|X_k| > r) = \Prb^!(\Phi(D_r) \le k-1)$.
Since $\Exp^!\Phi(D_r) \ge \lambda\,\pi\,r^2 - 1$ from
\eqref{eq:Palmmean}, applying this to \eqref{eq:C-H} yields
\[
  \Prb^!\bigl(
    \Phi(D_r) \le \epsilon\,( \lambda\,\pi\,r^2 - 1)
  \bigr)
  \le \ee^{c_\epsilon}\,\ee^{-c_\epsilon\,\lambda\,\pi\,r^2}\!.
\]
Hence, for any $\epsilon\in(0,1)$ and $k\in\N$, we can take $r>0$
satisfying $\epsilon\,(\lambda\,\pi\,r^2 - 1)\ge k-1$, which implies
\eqref{eq:bound1}.

\paragraph*{\it Proof of (ii)}
We here derive an upper bound of $\Prb^0(R(o)>r)$ by using Foss \&
Zuyev's seven petals~\cite{FossZuye96}, which are considered to obtain
an upper bound of the tail probability for the circumscribed radius of
the typical Poisson-Voronoi cell.
Consider a collection of seven disks of common radii $r$ centered at
the points~$(r, 2\pi k/7)$, $k=0,1,\ldots,6$, in polar coordinates.
The petal~$0$ is given as the intersection of two circles centered at
$(r,0)$, $(r, 2\pi/7)$ and the angular domain between the rays
$\phi=0$ and $\phi=2\pi/7$.
The petal~$k$ is the rotation copy of petal~$0$ by angle $2\pi k/7$
for $k=1,2,\ldots,6$ (see Figure~\ref{fig:petals}).
Let $\mathcal{P}_{k,r}$, $k=0,1,\ldots,6$, denote the set formed by
petal~$k$ on the complex plane~$\C$.
Then, according to the discussion in the proof of Lemma 1 of
\cite{FossZuye96},
\begin{align}\label{eq:FoZu96}
  \Prb^0(R(o) > r)
  &\le \Prb^!\Bigl(
         \bigcup_{k=0}^6\{\Phi(\mathcal{P}_{k,r})=0\}
       \Bigr)
  \nonumber\\
  &\le 7\,\Prb^!\bigl(\Phi(\mathcal{P}_{0,r})=0\bigr),
\end{align}
where the second inequality follows from the isotropy of the Palm
version of $\Phi$.
Now, we can apply the bound~\eqref{eq:C-H} with $\epsilon=0$ and we have
\begin{equation}\label{eq:C-H-petal}
  \Prb^!\big(
    \Phi(\mathcal{P}_{0,r}) = 0
  \bigr)
  \le \ee^{-c_0\,\Exp^!\Phi(\mathcal{P}_{0,r})}.
\end{equation}
Hence, \eqref{eq:bound2} holds since $\Exp^!\Phi(\mathcal{P}_{0,r})
\ge \lambda\,\mu(\mathcal{P}_{0,r}) - 1$ and $\mu(\mathcal{P}_{0,r}) =
2\,r^2\,\bigl(\pi/7 + \sin(\pi/7)\,\cos(3\,\pi/7)\bigr)$.
\end{IEEEproof}

\begin{remark}
We can take $c_0$ in \eqref{eq:C-H-petal} equal to the unity since
determinantal point processes are weakly sub-Poisson (in particular,
due to the $\nu$-weakly sub-Poisson property) (see \cite{BlasYoge14}
for details).
\end{remark}

\begin{remark}
When the kernel~$K$ of the determinantal point process is explicitly
specified, it may be possible to obtain a tighter upper bound on the
tail probability of the circumscribed radius of the typical Voronoi
cell.
For example, the case of the Ginibre point process is given by the
following proposition.
\end{remark}

\begin{proposition}\label{prp:Ginibre-radius}
For the Ginibre point process (with intensity~$\pi^{-1}$), 
the circumscribed radius for the typical Voronoi cell
$\mathcal{C}(o)$ satisfies
\begin{equation}\label{eq:covradius}
  \Prb^0(R(o) > r)
  \le 7\,\ee^{-(u(r)\vee v(r))},
\end{equation}
where
\begin{align*}
  u(r)
  &= \frac{1}{7}\,
     \Bigl(
       4r^2\,\cos^2\frac{2\pi}{7}
       + \exp\Bigl(-4 r^2\,\cos^2\frac{2\pi}{7}\Bigr) - 1
     \Bigr),
  \\
  v(r)
  &= \frac{2 r^2}{\pi}\,
     \Bigl(
       \frac{\pi}{7} + \sin\frac{\pi}{7}\,\cos\frac{3 \pi}{7}
     \Bigr)
  \\
  &\quad\mbox{}
     + \frac{1}{7}\,
       \Bigl(
         \exp\Bigl(-4 r^2\,\cos^2\frac{\pi}{7}\Bigr) - 1
       \Bigr).
\end{align*}
\end{proposition}

\begin{IEEEproof}
The kernel of the Ginibre point process is given by
\[
  K(z, w) =
  \frac{1}{\pi}\,\ee^{-(|z|^2+|w|^2)/2}\,\ee^{z\,\Bar{w}},
  \quad z, w\in\C,
\]
with respect to the Lebesgue measure (see, e.g., \cite{ShirTaka03}).
Thus, the intensity function~\eqref{eq:1-corr} of the (reduced) Palm
version reduces to
\begin{equation}\label{eq:Ginibre-Palm-intensity}
  \rho_1^0(z)
  = \frac{1}{\pi}\,\bigl(1 -  \ee^{-|z|^2}\bigr),
  \quad z\in\C.
\end{equation}
Now, we obtain two lower bound of $\Exp^!\Phi(\mathcal{P}_{0,r})$ as
follows.
Let $\mathcal{S}_\eta$ denote the circular sector centered at the
origin with radius~$\eta$ and angular domain between $\phi=0$ and
$\phi=2\,\pi/7$.
When we take $\eta_1=2r\,\cos(2\pi/7)$ and $\eta_2=2r\,\cos(\pi/7)$,
we have
$\mathcal{S}_{\eta_1}\subset\mathcal{P}_{0,r}\subset\mathcal{S}_{\eta_2}$.
Therefore, applying \eqref{eq:Ginibre-Palm-intensity}, we have the
first lower bound;
\begin{align*}
  \Exp^!\Phi(\mathcal{P}_{0,r})
 &\ge \Exp^!\Phi(\mathcal{S}_{\eta_1})
  = \int_{\mathcal{S}_{\eta_1}} \rho^0_1(z)\,\dd z
 \\
 &= \frac{1}{7}\,({\eta_1}^2 + \ee^{-{\eta_1}^2} - 1) = u(r).
\end{align*}
The second lower bound is given by
\begin{align*}
  \Exp^!\Phi(\mathcal{P}_{0,r})
 &=  \int_{\mathcal{P}_{0,r}}\rho_1^0(z)\,\dd z
  \\
 &\ge \frac{1}{\pi}\,
      \Bigl(
        \mu(\mathcal{P}_{r,0}) -
        \int_{\mathcal{S}_{\eta_2}}\ee^{-|z|^2}\,\dd z
      \Bigr) = v(r).
\end{align*}
Hence, we have \eqref{eq:covradius} from \eqref{eq:FoZu96} and
\eqref{eq:C-H-petal} with $c_0=1$.
\end{IEEEproof}

Indeed, there is $r_*\approx0.5276\cdots$ such that $u(r)>v(r)$ for $r<r_*$
and $u(r)<v(r)$ for $r>r_*$.

We are now ready to give the tail asymptotics of the SIR distribution
when the BSs are deployed according to the Ginibre point process.

\begin{corollary}\label{cor:Ginibre}
Suppose that $\Phi=\{X_i\}_{i\in\N}$ is the Ginibre point process.
When the propagation effect sequence~$\{H_i\}_{i\in\N}$ satisfies the
condition~(\ref{condB}) of Theorem~\ref{thm:general}, we have
\begin{align}\label{eq:GiniAsym}
 &\lim_{\theta\to\infty}\theta^{1/\beta}\,
    \Prb(\SIR_o > \theta)
 \nonumber\\
 &= \frac{\Exp({H_1}^{1/\beta})}{\Gamma(1+1/\beta)}
    \int_0^\infty
      \prod_{i=1}^\infty
        \frac{1}{i!}
        \int_0^\infty
          \ee^{-u}\,u^i\,
          \Lpl_H\Bigl(\Bigl(\frac{t}{u}\Bigr)^\beta\Bigr)\,
        \dd u\,
    \dd t.
\end{align}
Furthermore, when $H_i\sim\Gam(m,\:1/m)$ (Nakagami-$m$ fading without
shadowing),
\begin{align}\label{eq:GiniNakaAsym}
 &\lim_{\theta\to\infty}\theta^{1/\beta}\,
    \Prb(\SIR_o > \theta)
 \nonumber\\
 &= \frac{\beta}{B(m,\:1/\beta)}
    \int_0^\infty
      \prod_{i=1}^\infty
        \frac{1}{i!}
        \int_0^\infty
          \frac{\ee^{-u}\,u^i}
               {\bigl(1 + (v/u)^\beta\bigr)^m}\,
        \dd u\,
    \dd v,
\end{align}
where $B$ denotes the Beta function.
\end{corollary}

For the proof of Corollary~\ref{cor:Ginibre}, we use the following
proposition which is a consequence of \cite{Gold10} and \cite{Kost92}.

\begin{proposition}\label{prp:Kostlan}
Let $X_i$, $i\in\N$, denote the points of the reduced Palm version of
the Ginibre point process.
Then, the set $\{|X_i|\}_{i\in\N}$ has the same distribution as
$\{\sqrt{Y_i}\}_{i\in\N}$, where $Y_i$, $i\in\N$, are mutually
independent and $Y_i\sim\Gam(i+1, 1)$ for each $i\in\N$.
\end{proposition}

\begin{IEEEproof}[Proof of Corollary~\ref{cor:Ginibre}]
For the Ginibre point process, we can see from
Lemma~\ref{lem:determinantal} (or
Proposition~\ref{prp:Ginibre-radius}) that $|X_k|$, $k\in\N$, and
$R(o)$ have any order of moments with respect to the Palm probability
$\Prb^0$; that is, the condition~(\ref{condA}) of
Theorem~\ref{thm:general} is fulfilled.
Thus, applying the identity $x^{-1/\beta} =
\Gamma(1/\beta)^{-1}\int_0^\infty \ee^{-x\,s}\, s^{-1+1/\beta}\,\dd s$
to the right-hand side of \eqref{eq:GH15Thm4}, we have
\begin{align*}
 &\Exp^0\biggl[
    \biggl(
      \sum_{i\in\N}\frac{H_i}{|X_i|^{2\,\beta}}
    \biggr)^{-1/\beta}
  \biggr]
 \\
  &= \frac{1}{\Gamma(1/\beta)}
     \int_0^\infty
       s^{-1+1/\beta}\,
       \Exp^0\biggl[
         \exp\biggl(
           -s \sum_{i\in\N}
             \frac{H_i}{|X_i|^{2\,\beta}}
         \biggr)
       \biggr]\,
     \dd s
  \\
  &= \frac{1}{\Gamma(1/\beta)}
     \int_0^\infty
       s^{-1+1/\beta}\,
       \Exp^0\biggl[
         \prod_{i\in\N}
           \Lpl_H\biggl(
             \frac{s}{|X_i|^{2\,\beta}}
           \biggr)
       \biggr]\,
     \dd s
  \\
  &= \frac{1}{\Gamma(1+1/\beta)}
     \int_0^\infty
       \Exp^0\biggl[
         \prod_{i\in\N}
           \Lpl_H\biggl(
             \biggl(\frac{t}{|X_i|^2}\biggr)^\beta
           \biggr)
       \biggr]\,
     \dd t,
\end{align*}
where the last equality follows by substituting $t=s^{1/\beta}$.
Here, Proposition~\ref{prp:Kostlan} states that
$\{|X_i|^2\}_{i\in\N}\eqd\{Y_i\}_{i\in\N}$ under $\Prb^0$ with
$Y_i\sim\Gam(i+1,\: 1)$, so that, applying the density function of
$\Gam(i+1,\: 1)$, $i\in\N$, we have \eqref{eq:GiniAsym}.

When $H_i\sim\Gam(m,\:1/m)$, then $\Lpl_H(s)=(1+s/m)^{-m}$ and
$\Exp({H_1}^{1/\beta}) =
\Gamma(m+1/\beta)/\bigl(m^{1/\beta}\,(m-1)!\bigr)$.
Thus, applying these to the right-hand side of \eqref{eq:GiniAsym},
\begin{align*}
  \eqref{eq:GiniAsym}
  &= \frac{\Gamma(m+1/\beta)}
         {\Gamma(1+1/\beta)\,m^{1/\beta}\,(m-1)!}
 \\
  &\quad\mbox{}\times
     \int_0^\infty
       \prod_{i=1}^\infty
         \frac{1}{i!}
         \int_0^\infty
           \frac{\ee^{-u}\,u^i}
                {\bigl(1+m^{-1}\,(t/u)^\beta\bigr)^m}\,
         \dd u\,
     \dd t.
\end{align*}
Finally, substituting $v=m^{-1/\beta}\,t$ and applying
$B(x,y)=\Gamma(x)\,\Gamma(y)/\Gamma(x+y)$, we obtain
\eqref{eq:GiniNakaAsym}.
%
\end{IEEEproof}

\begin{remark}
When $m=1$, \eqref{eq:GiniNakaAsym} reduces to the result of
\cite[Theorem~2]{MiyoShir14a}, which considers the Rayleigh fading.
\end{remark}

\subsection{A counterexample}

Finally, we give a simple counterexample that violates the
condition~(\ref{condA}) of Theorem~\ref{thm:general}.
Let $T$ denote a random variable with density function
$f_T(t)=(a-1)\,t^{-a}$, $t\ge1$, for $a\in(1,2)$.
Given a sample of $T$, we consider the mixed and randomly shifted
lattice $\Phi = (\Z\times T\,\Z) + U_T$, where $U_T$ denotes a
uniformly distributed random variable on $[0,1]\times[0,T]$.
The intensity $\lambda$ of $\Phi$ is then
$\lambda=\Exp(1/T) = (a-1)/a<\infty$.
For any nonnegative and measurable function~$f$, the definition of the
Palm probability implies
\begin{align*}
  \Exp^0f(T)
  &= \frac{1}{\lambda}\,
     \Exp\bigl(f(T)\,\Phi(I)\bigr)
  \\
  &= \frac{1}{\lambda}\,
     \Exp\bigl(f(T)\,\Exp(\Phi(I)\mid T)\bigr)
   = \frac{1}{\lambda}\,\Exp\Bigl(\frac{f(T)}{T}\Bigr),
\end{align*}
where $I=[0,1]\times[0,1]$.
Hence, applying $R(o)^2 = (1+T^2)/4$ to the above,
we have
\begin{align*}
  \Exp^0\bigl(R(o)^2\bigr)
  = \frac{1}{4\,\lambda}\,
    \Exp\Bigl(
      \frac{1}{T}+T
    \Bigr)
  = \frac{1}{4\lambda}\,\bigl(\lambda + \Exp(T)\bigr)=\infty.
\end{align*}

\section{Concluding remark}

More other examples will be investigated in the extended version of
the paper.


\section*{Acknowledgment}

The first author wishes to thank Radha Krishna Ganti for directing his
interest to the SIR asymptotics in cellular network models based on
general stationary point processes.
The first author's work was supported by the Japan Society for the
Promotion of Science (JSPS) Grant-in-Aid for Scientific Research (C)
25330023.
The second author's work was supported by JSPS Grant-in-Aid for
Scientific Research (B) 26287019.

\appendix
\begin{IEEEproof}[Proof of Lemma~\ref{lem:kernel}]
For a compact set $C\in\B(\C)$, let $K_C$ denote the restriction of
$K$ on $C$.
Let also $\kappa_{C,i}$ and $\varphi_{C,i}$, $i\in\N$, denote
respectively the eigenvalues of $K_C$ and the corresponding
orthonormal eigenfunctions; that is,
\begin{equation}\label{eq:orthonormal}
  \int_C
    \varphi_{C,i}(z)\,\Bar{\varphi_{C,j}(z)}\,
  \dd z
  = \begin{cases}
      1 &\quad\text{for $i=j$,}\\
      0 &\quad\text{for $i\ne j$.}
    \end{cases}%
\end{equation}
Then Mercer's theorem states that the following spectral expansion
holds (see, e.g., \cite{Merc09});
\begin{equation}\label{eq:Mercer}
  K_C(z,w)
  = \sum_{i=1}^\infty
      \kappa_{C,i}\,\varphi_{C,i}(z)\,\Bar{\varphi_{C,i}(w)},
  \quad z, w\in C.
\end{equation}
Note that $\kappa_{C,i}\in[0,1]$, $i\in\N$, under the assumption on
the kernel~$K$.
Thus we have
\begin{align*}
  \int_C|K(0,z)|^2\,\dd z
  &= \int_C|K_C(0,z)|^2\,\dd z
  \\
  &= \sum_{i=1}^\infty
       {\kappa_{C,i}}^2\,|\varphi_{C,i}(0)|^2
  \\
  &\le K_C(0,0) = K(0,0),
\end{align*}
where the second equality follows from \eqref{eq:orthonormal} and the
inequality holds since $\kappa_{C,i}\in[0,1]$, $i\in\N$, and
\eqref{eq:Mercer}.
Hence, letting $C\uparrow\C$, we obtain \eqref{eq:boundM}.
\end{IEEEproof}


\begin{thebibliography}{10}
\providecommand{\url}[1]{#1}
\csname url@samestyle\endcsname
\providecommand{\newblock}{\relax}
\providecommand{\bibinfo}[2]{#2}
\providecommand{\BIBentrySTDinterwordspacing}{\spaceskip=0pt\relax}
\providecommand{\BIBentryALTinterwordstretchfactor}{4}
\providecommand{\BIBentryALTinterwordspacing}{\spaceskip=\fontdimen2\font plus
\BIBentryALTinterwordstretchfactor\fontdimen3\font minus
  \fontdimen4\font\relax}
\providecommand{\BIBforeignlanguage}[2]{{%
\expandafter\ifx\csname l@#1\endcsname\relax
\typeout{** WARNING: IEEEtran.bst: No hyphenation pattern has been}%
\typeout{** loaded for the language `#1'. Using the pattern for}%
\typeout{** the default language instead.}%
\else
\language=\csname l@#1\endcsname
\fi
#2}}
\providecommand{\BIBdecl}{\relax}
\BIBdecl

\bibitem{BaccBlas09a}
F.~Baccelli and B.~B{\l}aszczyszyn, ``Stochastic geometry and wireless
  networks, {Volume~I:\ Theory},'' \emph{Foundations Trends~(R)
    Networking}, vol.~3, pp. 249--449, 2009.

\bibitem{Haen13}
M.~Haenggi, \emph{Stochastic Geometry for Wireless networks}.\hskip
  1em plus 0.5em minus 0.4em\relax Cambridge Univ.\ Press, 2013.

\bibitem{Mukh14}
S.~Mukherjee, \emph{Analytical Modeling of Heterogeneous Cellular
  Networks: Geometry, Coverage, and Capacity}.\hskip 1em plus 0.5em
  minus 0.4em\relax Cambridge Univ.\ Press, 2014.

\bibitem{AndrBaccGant11}
J.~G. Andrews, F.~Baccelli, and R.~K. Ganti, ``A tractable approach to
  coverage and rate in cellular networks,'' \emph{IEEE
    Trans.\ Commun.}, vol.~59, pp. 3122--3134, 2011.

\bibitem{MiyoShir14a}
N.~Miyoshi and T.~Shirai, ``A cellular network model with {Ginibre}
configured base stations,'' \emph{Adv.\ Appl.\ Probab.}, vol.~46, pp.
  832--845, 2014.

\bibitem{NagaMiyoShir14}
H.~Nagamatsu, N.~Miyoshi, and T.~Shirai, ``Pad\'{e} approximation for
  coverage probability in cellular networks,''
  \emph{SpaSWiN-WiOpt 2014}, pp. 693--700, 2014.

\bibitem{Haen14}
M.~Haenggi, ``The mean interference-to-signal ratio and its key role in
  cellular and amorphous networks,'' \emph{IEEE Trans.\ Wireless
    Commun.}, vol.~3, pp. 597--600, 2014.

\bibitem{GuoHaen15}
A.~Guo and M.~Haenggi, ``Asymptotic deployment gain: {A} simple approach to
  characterize the {SINR} distribution in general cellular networks,''
  \emph{IEEE Trans.\ Commun.}, vol.~63, pp. 962--976, 2015.

\bibitem{GantHaen15}
R.~K. Ganti and M.~Haenggi, ``Asymptotics and approximation of the {SIR}
  distribution in general cellular networks,''
  \emph{IEEE Trans.\ Wireless Commun.}, vol.~15, pp. 2130--2143, 2016.

\bibitem{MiyoShir15}
N.~Miyoshi and T.~Shirai, ``Downlink coverage probability in a
cellular network with {Ginibre} deployed base stations and
{Nakagami-$m$} fading channels,'' \emph{WiOpt 2015}, pp. 483--489,
2015.

\bibitem{Sliv62}
I.~M. Slivnyak, ``Some properties of stationary flows of homogeneous random
  events,'' \emph{Theory Probab.\ Appl.}, vol.~7, pp. 336--341, 1962.

\bibitem{Calk02}
P.~Calka, ``The distributions of the smallest disks containing the
  {Poisson-Voronoi} typical cell and the {Crofton} cell in the plane,''
  \emph{Adv.\ Appl.\ Probab.}, vol.~34, pp. 702--717, 2002.

\bibitem{HougKrisPereVira09}
J.~B. Hough, M.~Krishnapur, Y.~Peres, and B.~Vir\'ag, \emph{Zeros of Gaussian
  Analytic Functions and Determinantal Point Processes}.\hskip 1em plus 0.5em
  minus 0.4em\relax American Math.\ Soc., 2009.

\bibitem{ShirTaka03}
T.~Shirai and Y.~Takahashi, ``Random point fields associated with certain
  {Fredholm determinants~I:\ Fermion, Poisson and Boson processes},''
  \emph{J.\ Funct.\ Analysis}, vol. 205, pp. 414--463, 2003.


\bibitem{Cher52}
H.~Chernoff, ``A measure of asymptotic efficiency for tests of a hypothesis
  based on the sum of observations,'' \emph{Ann.\ Math.\ Statist.},
  vol.~23, pp. 493--507, 1952.

\bibitem{Hoef63}
W.~Hoeffding, ``Probability inequalities for sums of bounded random
  variables,'' \emph{J.\ American Statist.\ Assoc.}, vol.~58,
  pp. 13--30, 1963.

\bibitem{FossZuye96}
S.~G. Foss and S.~A. Zuyev, ``On a {Voronoi} aggregative process related to a
  bivariate {Poisson} process,'' \emph{Adv.\ Appl.\ Probab.},
  vol.~28, pp. 965--981, 1996.

\bibitem{BlasYoge14}
B.~B{\l}aszczyszyn and D.~Yogeshwaran, ``On comparison of clustering properties
  of point processes,'' \emph{Adv.\ Appl.\ Probab.}, vol.~46, pp.
  1--20, 2014.

\bibitem{Gold10}
A.~Goldman, ``The {Palm} measure and the {Voronoi} tessellation for the
  {Ginibre} process,'' \emph{Annals Appl.\ Probab.}, vol.~20, pp.
  90--128, 2010.

\bibitem{Kost92}
E.~Kostlan, ``On the spectra of {Gaussian} matrices,'' \emph{Linear
  Algebra Appl.}, vol. 162--164, pp. 385--388, 1992.

\bibitem{Merc09}
J.~Mercer, ``Functions of positive and negative type, and their connection with
  the theory of integral equations,'' \emph{Philos.\ Trans.\ Royal
    Soc.\ London: Ser.~A}, vol. 209, pp. 415--446, 1909.
\end{thebibliography}


\end{document}